\documentclass[useAMS,usenatbib]{mn2e}
\usepackage{epsf}
\title[Supernova constraints on alternative models to dark energy]{Supernova constraints on alternative models to dark energy}
\author[Y Gong and C.K. Duan]{Yungui Gong\thanks{E-mail:
gongyg@cqupt.edu.cn} and Chang-Kui Duan\thanks{E-mail:
duanck@cqupt.edu.cn}\\
Institute of Applied Physics and College of Electronic
Engineering, \\
Chongqing University of Posts and Communications,
Chongqing 400065, China}
\begin{document}

\maketitle

\begin{abstract}
The recent observations of type Ia supernovae suggest that the
universe is accelerating now and decelerated in the recent past.
This may be the evidence of the breakdown of the standard
Friedmann equation. The Friedmann equation $H^2\sim \rho$ is
modified to be a general form $H^2=g(\rho)$. Three models with
particular form of $g(\rho)$ are considered in detail. The
supernova data published by \citet{tonry}, \citet{ddsg} and
\citet{raknop03} are used to analyze the models. After the best
fit parameters are obtained, we then find out the transition
redshift $z_{\rm T}$ when the universe switched from the
deceleration phase to the acceleration phase.

\end{abstract}


\section{Introduction}
The type Ia supernova observations suggest that the expansion of
the universe is speeding up rather than slowing down
(\citealt{sp97},1999; \citealt{gpm98}; \citealt{rag98};
\citealt{tonry}; \citealt{raknop03}). The measurements of the
anisotropy of the cosmic microwave background favor a flat
universe (\citealt{pdb00}; \citealt{hanany}; \citealt{bennett03};
\citealt{DNSpergel}). The observation of type Ia supernova SN
1997ff at $z\sim 1.7$ also supports that the universe was in the
deceleration phase in the recent past \citep{agr}. The transition
from the deceleration phase to the acceleration phase happened
around the redshift $z_{\rm T}\sim 0.5$ (\citealt{mstagr};
\citealt{ddsg}). A form of matter with negative pressure widely
referred as dark energy is usually introduced to explain the
accelerating expansion. The simplest form of dark energy is the
cosmological constant with the equation of state
$p_\Lambda=-\rho_\Lambda$. The cosmological constant model can be
easily generalized to dynamical cosmological constant models, such
as the dark energy model with the equation of state $p_{\rm
Q}=\omega_{\rm Q}\rho_{\rm Q}$, where the constant $\omega_{\rm
Q}$ satisfies $-1\le \omega_{\rm Q}<-1/3$. If we remove the null
energy restriction $\omega_{\rm Q}\ge -1$ to allow supernegative
$\omega_{\rm Q}<-1$, then we get the phantom energy models
(\citealt{ajs03}; \citealt{carr02}; \citealt{kmbs}). More exotic
equation of state is also possible, such as the Chaplygin gas
model with the equation of state $p=-A/\rho$ and the generalized
Chaplygin gas model with the equation of state $p=-A/\rho^\alpha$
(\citealt{chaply}; \citealt{bntgg}; \citealt{bmcbo};
\citealt{cdff}; \citealt{amff}; \citealt{caljas}). In general, a
scalar field $Q$ that slowly evolves down its potential $V(Q)$
takes the role of a dynamical cosmological constant
(\citealt{quint}; \citealt{zlwl}; \citealt{pgfmj},1998;
\citealt{brpjep}; \citealt{ptw99}; \citealt{johri};
\citealt{rcbjd}; \citealt{johri02}; \citealt{lautm};
\citealt{ulamt}; \citealt{aasss}; \citealt{gong02},2004). The
scalar field $Q$ is also called the quintessence field. The energy
density of the quintessence field must remain very small compared
to that of radiation and matter at early epoches and evolves in a
way that it started to dominate the universe around the redshift
$0.5$. There are other forms of dark energy, like the tachyon
field (\citealt{tachyon}; \citealt{ptctr02},2003; \citealt{bagla};
\citealt{padmt03}; \citealt{ptct03}).

Although dark energy models are consistent with current
observations, the nature of dark energy is still a mystery.
Therefore it is possible that the observations show a sign of the
breakdown of the standard cosmology. Some alternative models to
dark energy models were proposed along this line of reasoning.
These models are motivated by brane cosmology (\citealt{dgp};
\citealt{dvali03a}; \citealt{chung}). In this scenario, our
universe is a three-brane embedded in a five dimensional
spacetime. The five dimensional action is
$$S_5={-1\over 2\kappa_5}\int d^5x\sqrt{G}\,\mathcal{R}+S_{\rm orb}+S_{\rm boundary}+S_{\rm GH},$$
where $\mathcal{R}$ is the Ricci scalar in five dimensions, $G$ is
the five dimensional metric determinant, $\kappa_5$ is the five
dimensional Newton's constant, $S_{\rm orb}$ is the orbifold
action, $S_{\rm boundary}$ represents the boundary action and
$S_{\rm GH}$ is the Gibbons-Hawking boundary terms. In these
models, the usual Friedmann equation $H^2=8\pi G\rho/3$ is
modified to a general form $H^2=g(\rho)$ and the universe is
composed of the ordinary matter only (\citealt{freese02};
\citealt{freese03}; \citealt{gondolo}; \citealt{sen03},2003b;
\citealt{zhu03},2003b,2003c; \citealt{zfh}; \citealt{wang};
\citealt{maltamaki}; \citealt{dev03}; \citealt{frith};
\citealt{gong03}; \citealt{gcd03}). One particular example is the
brane cosmology with $g(\rho)\sim \rho+\rho^2$. In order to retain
the success of the standard cosmology at early times, we require
that the modified cosmology recovers the standard cosmology at
early times. So $g(\rho)$ must satisfy $g(\rho)\sim \rho$ when
$\rho \gg \rho_0$, where $\rho_0$ is the current matter energy
density.

For a spatially flat, isotropic and homogeneous universe with both
an ordinary pressureless dust matter and a minimally coupled
scalar field $Q$ sources, the Friedmann equations are
\begin{eqnarray}
\label{cos1} H^2=\left({\dot{a}\over a}\right)^2={8\pi
G\over 3}(\rho_{\rm m}+\rho_{\rm Q}),\\
\label{cos2}
{\ddot{a}\over a}=-{4\pi G\over
3}(\rho_{\rm m}+\rho_{\rm Q}+3p_{\rm Q}),\\
\label{cos3} \dot{\rho_{\rm Q}}+3H(\rho_{\rm Q}+p_{\rm Q})=0,
\end{eqnarray}
where dot means derivative with respect to time, $\rho_{\rm
m}=\rho_{\rm m0}(a_0/a)^3$ is the matter energy density, a
subscript 0 means the value of the variable at present time,
$\rho_{\rm Q}=\dot{Q}^2/2+V(Q)$, $p_{\rm Q}=\dot{Q}^2/2-V(Q)$ and
$V(Q)$ is the potential of the quintessence field. The modified
Friedmann equations for a spatially flat universe are
\begin{eqnarray}
\label{cosa} H^2=H_0^2g(x),\\
\label{cosb} {\ddot{a}\over a}=H^2_0g(x)
-{3H^2_0x\over 2}g'(x)\left({\rho+p\over \rho}\right),\\
\label{cosc} \dot{\rho}+3H(\rho+p)=0,
\end{eqnarray}
where $x=8\pi G\rho/3H^2_0=x_0(1+z)^3$ during the matter dominated
epoch, $1+z=a_0/a$ is the redshift parameter, $g(x)=x+\cdots$ is a
general function of $x$ and $g'(x)=dg(x)/dx$. Note that the
universe did not start to accelerate when the other nonlinear
terms in $g(x)$ started to dominate. To recover the standard
cosmology at early times, we require that $g(x)\approx x$ when
$x\gg x_0$. For the matter dominated flat universe,
$\rho=\rho_{\rm m}$ and $p=p_{\rm m}=0$. Let $\Omega_{\rm m0}=8\pi
G\rho_0/3H^2_0$, then $x_0=\Omega_{\rm m0}$, $g(x_0)=1$ and
$x=\Omega_{\rm m0}(1+z)^3$ during the matter dominated era.

The luminosity distance $d_{\rm L}$ is defined as
\begin{eqnarray}
\label{lumin} d_{\rm L}(z)=a_0(1+z)\int^{t_0}_t {dt'\over
a(t')}\nonumber\\
={1+z\over H_0}\int^z_0 g^{-1/2}[\Omega_{\rm m0}(1+u)^3]du.
\end{eqnarray}
The apparent magnitude redshift relation is
\begin{eqnarray}
\label{magn}
 m(z)=M+5\log_{10}d_{\rm L}(z)
+25=\mathcal{M}+5\log_{10}\mathcal{D}_{\rm L}(z) \nonumber \\
=\mathcal{M}+5\log_{10}\left\{(1+z)\int^z_0 g^{-1/2}[\Omega_{\rm
m0}(1+u)^3]du\right\},
\end{eqnarray}
where $\mathcal{D}_{\rm L}(z)=H_0d_{\rm L}(z)$ is the
``Hubble-constant-free" luminosity distance, $M$ is the absolute
peak magnitude and $\mathcal{M}=M-5\log_{10}H_0+25$. $\mathcal{M}$
can be determined from the low redshift limit at where
$\mathcal{D}_{\rm L}(z)=z$. The parameters in our model are
determined by minimizing
\begin{equation}
\label{lrmin}
\chi^2=\sum_i{[m_{obs}(z_i)-m(z_i)]^2\over \sigma^2_i},
\end{equation}
where $\sigma_i$ is the total uncertainty in the observations. The
$\chi^2$-minimization procedure is based on MINUIT code. We use
the 54 supernova data with both the stretch correction and the
host-galaxy extinction correction, i.e., the fit 3 supernova data
by \cite{raknop03}, the 20 radio galaxy and 78 supernova data by
\cite{ddsg}, and the supernova data by \cite{tonry} to find the
best fit parameters. In the fit, the range of parameter space for
$\mathcal{M}$ is $\mathcal{M}=[-3.9,\ 3.2]$, the range of
parameter space for $\Omega_{\rm m0}$ is $\Omega_{\rm m0}=[0,\
4]$.

The transition from deceleration to acceleration happens when the
deceleration parameter $q=-\ddot{a}/aH^2=0$. From equations
(\ref{cosa}) and (\ref{cosb}), we have
\begin{equation}
\label{trans} g[\Omega_{\rm m0}(1+z_{\rm T})^3]={3\over
2}\Omega_{\rm m0}(1+z_{\rm T})^3g'[\Omega_{\rm m0}(1+z_{\rm
T})^3],
\end{equation}
\begin{equation}
 \label{qparam} q_0={3\over 2}\Omega_{\rm m0}g'(\Omega_{\rm m0})-1.
\end{equation}
To compare the modified model with the dark energy model, we make
the following identification
\begin{equation}
\label{darkom} \omega_{Q}={xg'(x)-g(x)\over g(x)-x}.
\end{equation}

\section{Chaplygin gas Model}
The chaplygin gas model $p=-A/\rho^\alpha$ in the framework of
alternative model to dark energy is
$$g(x)=x+\Omega_{\rm Q0}[A_s+(1-A_s)(x/\Omega_{\rm m0})^\beta]^{1/\beta},$$
where $\Omega_{\rm Q0}=1-\Omega_{\rm m0}$, $\beta=1+\alpha$ and
$A_s=(8\pi G/3H^2_0\Omega_{\rm Q0})^\beta A$. The $\alpha=1$ model
is motivated by a $d$-brane in $d+2$ spacetime. Since
$g'(x)=1+\Omega_{\rm Q0}(1-A_s)[A_s+(1-A_s)(x/\Omega_{\rm
m0})^\beta]^{1/\beta-1}(x/\Omega_{\rm m0})^\beta$, so
\begin{eqnarray}
{\Omega_{\rm m0}\over 2\Omega_{\rm
Q0}}(1+z_{q=0})^3[A_s+(1-A_s)(1+z_{q=0})^{3\beta}]^{1-1/\beta}\nonumber\\
=A_s-{1\over 2}(1-A_s)(1+z_{q=0})^{3\beta},
\end{eqnarray}
\begin{equation}
q_0={1\over 2}-{3\over 2}A_s(1-\Omega_{\rm m0}),
\end{equation}
$q_0<0$ gives that $A_s>(1-\Omega_{\rm m0})^{-1}/3$. To retain the
success of the standard model at early epoches, we require
$g(x)\approx x$ when $x\gg 1$. In other words, we require $A_s\sim
1$. Therefore, we have the following constraints
\begin{eqnarray}
\label{chpl1}
(1-\Omega_{\rm m0})^{-1}/3<A_s<1,\\
\label{chpl2} A_s\sim 1.
\end{eqnarray}
The best fits to the 54 supernovae by \cite{raknop03} are
$\Omega_{\rm m0}=[0,\ 0.44]$ centered at almost zero, $A_s=[0.99,\
1]$ centered at almost one and $\beta=[1,\ 32.3]$ centered at 22.0
with $\chi^2=43.9$. The best fits to the 98 radio galaxy and
supernova data compiled by \cite{ddsg} are $\Omega_{\rm m0}=[0,\
0.27]$ centered at 0.26, $A_s=[0.65,\ 1]$ centered at 0.97 and
$\beta=1$ with $\chi^2=87.8$. The best fits to the 172 supernovae
with redshift $z>0.01$ and $A_v<0.5$ mag \citep{tonry} are
$\Omega_{\rm m0}=[0,\ 0.3]$ centered at 0.1, $A_s=[0.99,\ 1]$
centered at almost one and $\beta=[1.5,\ 23.7]$ centered at 16.1
with $\chi^2=169.5$. The best fits to the 194 supernovae by
\cite{tonry} and \cite{barris} are $\Omega_{\rm m0}=[0,\ 0.36]$
centered at 0.19, $A_s=[0.99,\ 1]$ centered at almost one and
$\beta=[1.3,\ 25.2]$ centered at 14.9 with $\chi^2=195.1$. The
best fits to all the data combined are $\Omega_{\rm m0}=[0,\
0.34]$ centered at 0.21, $A_s=[0.99,\ 1]$ centered at almost one
and $\beta=[1.0,\ 21.2]$ centered at 13.7 with $\chi^2=331.3$.
From the above results, we conclude that the generalized Chaplygin
gas model tends to be the $\Lambda$ model.

\section{Generalized Cardassian model}
The model is
$$g(x)=x[1+Bx^{\alpha(n-1)}]^{1/\alpha},$$
where $B=(\Omega_{\rm m0}^{-\alpha}-1)/\Omega_{\rm
m0}^{\alpha(n-1)}$, $\alpha>0$ and $n<1-1/3(1-\Omega^\alpha_{\rm
m0})$. When $n=0$, $g(x)=B^{1/\alpha}(1+x^\alpha/B)^{1/\alpha}$
which is the case studied by \cite{freese03}. For the special case
$\alpha=1$ and $n=0$, $g(x)=x+B$ which is the standard cosmology
with a cosmological constant. From a purely phenomenological point
of view we may think that gravity is modified in such a way that
acceleration kicks in when the energy density approaches a certain
value \citep{carroll}. The model is motivated from a three-brane
located at the $Z_2$ symmetry fixed plane of a five dimensional
spacetime. Chung and Freese showed that if one parametrizes the
Hubble rate in terms of the brane energy density, then Cardassian
model is derived with suitable choice of the five dimensional
energy momentum tensor \citep{chung}. The generalized Cardassian
model gives
\begin{eqnarray}
\label{modb}
g'(x)&=&[1+Bx^{\alpha(n-1)}]^{1/\alpha}\nonumber\\
&&+(n-1)Bx^{\alpha(n-1)}[1+Bx^{\alpha(n-1)}]^{1/\alpha-1},
\end{eqnarray}
Combining equation (\ref{modb}) with equations (\ref{trans}) and
(\ref{qparam}), we get
\begin{eqnarray}
1+z_{\rm T}=[(\Omega_{\rm m0}^{-\alpha}-1)(2-3n)]^{1/3\alpha(1-n)},\\
q_0={1\over 2}+{3\over 2}(n-1)(1-\Omega_{\rm m0}^\alpha).
\end{eqnarray}
If we think the generalized Cardassian model as ordinary Freidmann
universe composed of matter and dark energy, we can identify the
following relationship for the parameters in the Cardassian and
quintessence models
$$\omega_{\rm Q0}={(n-1)(1-\Omega_{\rm m0}^\alpha)\over 1-\Omega_{\rm m0}}.$$

There are four parameters in the fits: the mass density
$\Omega_{\rm m0}$, the parameters $n$ and $\alpha$, as well as the
nuisance parameter $\mathcal{M}$. The range of parameter space
explored is: $n=[-10,\ 0.66]$ and $\alpha=(0,\ 10^4]$. The best
fits to the supernova data (\citealt{ddsg}; \citealt{tonry};
\citealt{raknop03}) generally give very large $\alpha>100$, so
$B\approx \Omega_{\rm m0}^{-\alpha n}$ and the transition redshift
is weakly dependent on $\alpha$. Furthermore, $\chi^2$ changes
very little when $\alpha$ changes over a fairly large range. In
other words, the generalized Cardassian model differs little from
the Cardassian model with $\alpha=1$. So we will discuss the
Cardassian model in more detail below.

\subsection{Cardassian Model}
\begin{figure}
\begin{center}
\epsfxsize=3.3in \epsffile{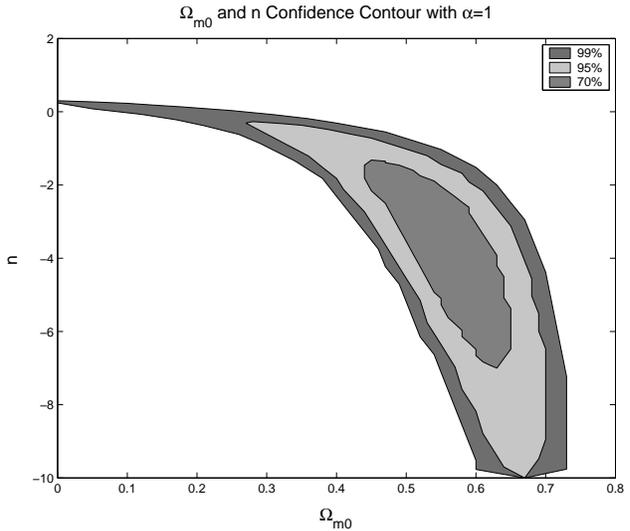}
\end{center}
\vspace{-0.2in} \caption{The 70\%, 95\% and 99\% confidence
contours of $\Omega_{\rm m0}$ and $n$ in Cardassian model from the
54 supernovae given by Knop et al. (2003)} \label{knopcont}
\end{figure}

The Cardassian model is the special case $\alpha=1$ of the
generalized Cardassian model. This model is equivalent to the dark
energy model with a constant equation of state $p_{\rm
Q}=\omega_{\rm Q}\rho_{\rm Q}$ in the sense of dynamical
evolution. The equivalence is provided by $n=1+\omega_{\rm Q0}$
and the equivalent dark energy potential is $V(Q)=A[\sinh
k(Q/\alpha+C)]^{-\alpha}$ with $\alpha=-2-2/(n-1)$. The best fits
to the 54 supernovae by \cite{raknop03} are $\Omega_{\rm
m0}=0.56^{+0.09}_{-0.12}$, $n=-3.6^{+2.2}_{-3.4}$ and
$\chi^2=43.73$. The $\Omega_{\rm m0}$ and $n$ contour plot is
shown in figure \ref{knopcont}.

\begin{figure}
\begin{center}
\epsfxsize=3.3in \epsffile{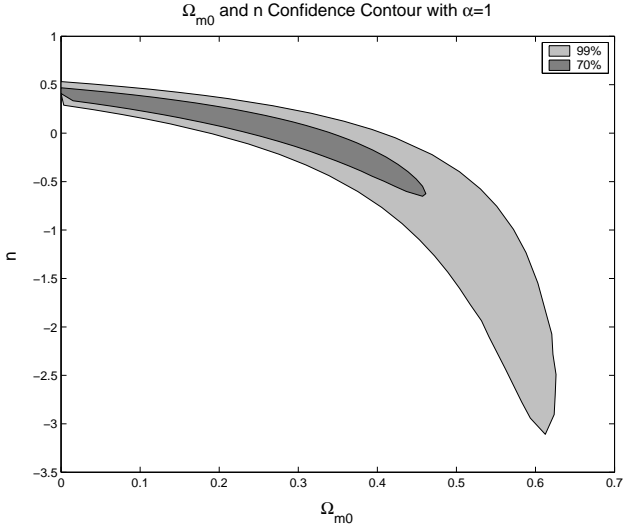}
\end{center}
\vspace{-0.2in} \caption{The 70\% and 99\% confidence contours of
$\Omega_{\rm m0}$ and $n$ in Cardassian model from 20 radio
galaxies and 78 supernovae compiled by Daly \& Djorgovski (2003)}
\label{dalycont}
\end{figure}

The best fits to the 98 radio galaxy and supernova data compiled
by \cite{ddsg} are $\Omega_{\rm m0}=0.14^{+0.32}_{-0.14}$,
$n=0.26^{+0.21}_{-0.91}$ and $\chi^2=87.45$. The contour plot is
shown in figure \ref{dalycont}.

\begin{figure}
\begin{center}
\epsfxsize=3.3in \epsffile{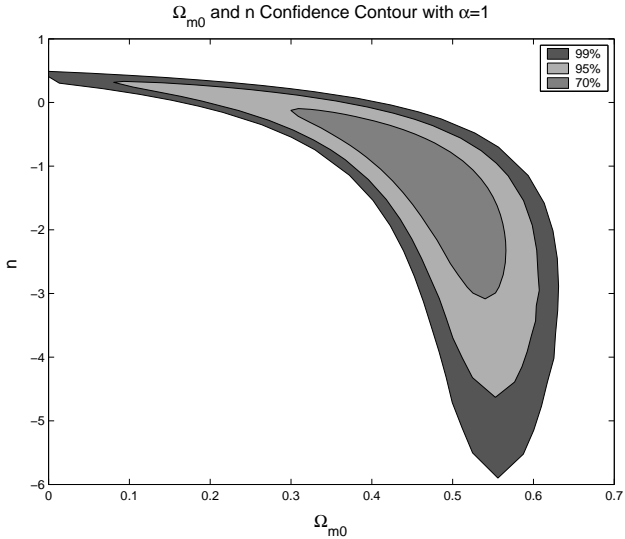}
\end{center}
\vspace{-0.2in} \caption{The 70\%, 95\% and 99\% confidence
contours of $\Omega_{\rm m0}$ and $n$ in Cardassian model from the
172 supernovae with $z>0.01$ and $A_v<0.5$ mag listed in Tonry et
al. (2003)} \label{frithcont2}
\end{figure}

The best fits to the 172 supernovae with redshift $z>0.01$ and
$A_v<0.5$ mag \citep{tonry} are $\Omega_{\rm
m0}=0.48^{+0.09}_{-0.18}$, $n=-1.2^{+1.1}_{-1.9}$ and
$\chi^2=171.4$. The best fits to the 194 supernovae by
\cite{tonry} and \cite{barris} are $\Omega_{\rm
m0}=0.51^{+0.08}_{-0.16}$, $n=-1.2^{+1.1}_{-1.9}$ and
$\chi^2=196.7$. The contour plot is shown is figure
\ref{frithcont2}. The plot agrees well with the figure 13 in
\cite{tonry} and the figure 6 in \cite{frith}.

\begin{figure}
\begin{center}
\epsfxsize=3.3in \epsffile{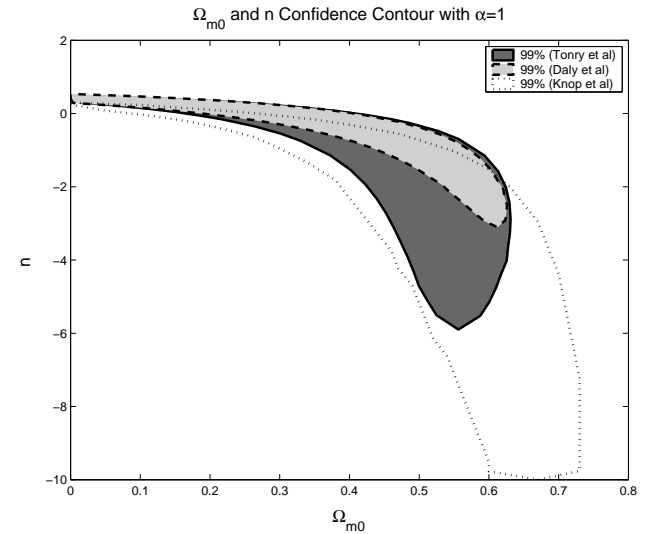}
\end{center}
\vspace{-0.2in} \caption{The 99\% confidence contours of
$\Omega_{\rm m0}$ and $n$ in Cardassian model from the sample data
in Tonry et al. (2003), Daly \& Djorgovski (2003) and Knop et al.
(2003)} \label{cont99}
\end{figure}

The above results are summarized in table \ref{tab1}. Combining
the above results, we find that $\Omega_{\rm m0}=[0,\ 0.61]$
centered at 0.45 and $n=[-3.1,\ 0.5]$ centered at $-0.6$ at 99\%
confidence level. The 99\% contour plot is shown in figure
\ref{cont99}. Take $\Omega_{\rm m0}=0.3$, we get $z_{\rm T}=0.35$,
$q_0=-3.1$ and $\omega_{\rm Q0}=-3.44$ when $n=-2.44$; $z_{\rm
T}=0.57$, $q_0=-0.24$ and $\omega_{\rm Q0}=-0.7$ when $n=0.3$.
These results are consistent with those obtained by Zhu \&
Fujimoto (2003a,b,c), \cite{zfh} and \cite{sen03b}.
\begin{table}
\caption{Best fits to Cardassian model} \label{tab1}
\begin{tabular}{lcccccc}
  \hline
  Fit &\multicolumn{2}{c}{$\Omega_{\rm m0}$}& \multicolumn{2}{c}{$n$}&$\chi^2$ \\
  \cline{2-5}
\#&70\%& 99\% & 70\% & 99\%&
  \\\hline
  1&$0.56^{+0.09}_{-0.12}$ & $0.56^{+0.17}_{-0.56}$&$-3.6^{+2.2}_{-3.4}$&$-3.6^{+3.9}_{-6.6}$&43.73 \\\hline
  2&$0.14^{+0.32}_{-0.14}$ & $0.14^{+0.48}_{-0.14}$&$0.26^{+0.21}_{-0.91}$&$0.26^{+0.27}_{-3.37}$&87.45 \\\hline
 3&$0.48^{+0.09}_{-0.18}$ & $0.48^{+0.15}_{-0.48}$&$-1.2^{+1.1}_{-1.9}$&$-1.2^{+1.7}_{-4.7}$&171.4 \\
  \hline
  4&$0.51^{+0.08}_{-0.16}$ & $0.51^{+0.14}_{-0.51}$&$-1.2^{+1.1}_{-1.9}$&$-1.2^{+1.8}_{-4.9}$&196.7 \\\hline
  5&$0.45^{+0.10}_{-0.19}$ & $0.45^{+0.16}_{-0.45}$&$-0.6^{+0.7}_{-1.1}$&$-0.6^{+1.1}_{-2.5}$&332.0 \\\hline
\end{tabular}
Fit 1 is the fit to the 54 supernova data from \cite{raknop03},
fit 2 is the fit to the 98 data points from \cite{ddsg}, fit 3 is
the fit to the 172 supernova data from \cite{tonry}, fit 4 is the
fit to the 194 supernova data from \cite{tonry} and \cite{barris}
and fit 5 is the fit to the above data combined.
\end{table}

\section{Model 3}
The last model we would like to consider is
$g(x)=[a+\sqrt{a^2+x}]^2$ (\citealt{Deffayet02a};
\citealt{dvali03a}), where $a=(1-\Omega_{\rm m0})/2$. This model
arises from the brane world theory by \cite{dgp} in which gravity
appears four dimensional at short distances while modified at
large distances. For this model, we find that the equivalent dark
energy equation of state parameter $\omega_{\rm Q0}$, $q_0$ and
the transition redshift $z_{\rm T}$ from decelerated expansion to
accelerated expansion are
\begin{eqnarray}
\omega_{\rm Q0}&=&-1/(1+\Omega_{\rm m0}),\\
q_0&=&{2\Omega_{\rm m0}-1\over 1+\Omega_{\rm m0}},\\
1+z_{\rm T}&=&\left[{2(1-\Omega_{\rm m0})^2\over \Omega_{\rm
m0}}\right]^{1/3}.
\end{eqnarray}
Applying the 54 supernova data with host-galaxy extinction
correction \citep{raknop03}, we find that $\Omega_{\rm
m0}=0.19^{+0.07}_{-0.05}$ and $\chi^2=45.71$. The 20 radio galaxy
and the 78 supernova data \citep{ddsg} give the best fit
$\Omega_{\rm m0}=0.18\pm 0.03$ and $\chi^2=87.6$. The best fit
from the 172 supernovae with redshift $z>0.01$ and $A_v<0.5$ mag
is $\Omega_{\rm m0}=0.17^{+0.04}_{-0.03}$ and $\chi^2=175.2$. If
we use the 194 supernovae given by \cite{tonry} and \cite{barris},
we find that $\Omega_{\rm m0}=0.22^{+0.04}_{-0.03}$ and
$\chi^2=200.6$. The above results are summarized in table
\ref{tab2}.
\begin{table}
\caption{Best fits to model 3} \label{tab2}
\begin{tabular}{rcccccc}
  \hline
  Data& \# &\multicolumn{2}{|c|}{$\Omega_{\rm m0}$} & $\chi^2$&$z_{\rm T}$&$\omega_{\rm Q0}$ \\
  \cline{3-4}
  source &  & $1\sigma$ & $3\sigma$ &  \\ \hline
  Knop & 54 & $0.19^{+0.07}_{-0.05}$ & $0.19^{+0.2}_{-0.12}$ & 45.71&0.90&$-0.84$ \\\hline
  Daly & 98 & $0.18\pm0.03$ & $0.18^{+0.1}_{-0.07}$ & 87.6&0.95&$-0.85$ \\\hline
  Tonry & 172 & $0.17^{+0.04}_{-0.03}$ & $0.17^{+0.14}_{-0.09}$ &
  175.2&1.0&$-0.85$
  \\\hline
  Barris & 194 & $0.22^{+0.04}_{-0.03}$ & $0.22^{+0.13}_{-0.09}$ & 200.6&0.77&$-0.82$ \\\hline
  Combined&346&$0.20\pm
  0.02$&$0.20^{+0.07}_{-0.06}$&334.6&0.86&$-0.83$ \\\hline
\end{tabular}
\end{table}
Combining the above results, we get $\Omega_{\rm m0}=0.20\pm 0.02$
at 1$\sigma$ level or $\Omega_{\rm m0}=0.20^{+0.07}_{-.0.06}$ at
3$\sigma$ level. If we take $\Omega_{\rm m0}=0.15$, then we have
$q_0=-0.61$ and $z_{\rm T}=1.13$. If we take $\Omega_{\rm
m0}=0.21$, then we have $q_0=-0.48$ and $z_{\rm T}=0.81$. If we
take $\Omega_{\rm m0}=0.28$, then we have $q_0=-0.34$ and $z_{\rm
T}=0.55$. These results are consistent with those obtained by
\cite{Deffayet02a}.

\section{Discussions and Conclusions}
A general function $g(x)$ of the ordinary matter density was used
to explain the current accelerating expansion of the universe. In
this model, no exotic matter form is needed. This approach is
equivalent to dark energy model building approach in the sense of
dynamical evolution of the universe because we can map the
modified part of energy density to dark energy. The function
$g(x)$ satisfies the following conditions: (1) $g(x_0)=1$; (2)
$g(x)\approx x$ when $z\gg 1$; (3) $g(x_0)>3x_0g'(x_0)/2$, where
$x=\Omega_{\rm m0}(1+z)^3+\Omega_{\rm r0}(1+z)^4$. The generalized
Chaplygin gas model tends to take $A_s=1$ which becomes the dark
energy model with a cosmological constant. Therefore the
generalized Chaplygin gas model is disfavored in the framework of
alternative models although it is a viable dark energy model.
Unlike the gravitational lensing constraint \citep{dev03}, the
supernova data do not provide tight constraint on the generalized
Cardassian model. A fairly large range of parameters from the
generalized Cardassian model are consistent with the supernova
data. For the Cardassian model, the supernova data give
$\Omega_{\rm m0}=[0,\ 0.62]$ and $n=[-3.11,\ 0.3]$ at the 99\%
confidence level. If we have better constraint on the transition
redshift $z_{\rm T}$, then we will be able to distinguish the
Cardassian model from the $\Lambda$ model because the $\Lambda$
model is the special case $n=0$. For the model 3, we find that
$\Omega_{\rm m0}=[0.12,\ 0.28]$ with 99.7\% confidence. Only the
upper limit gives $z_{\rm T}\sim 0.5$.

\section*{acknowledgments} The author Gong would like to thank Tonry
for pointing out the program he used to compute $\chi^2$ in his
webpage, the author Gong is also grateful to the discussion with
Frith. The work is supported by Chongqing University of Post and
Telecommunication under grants A2003-54 and A2004-05.

\end{document}